\documentclass[11pt]{article} 
\usepackage{amsmath}
\usepackage[utf8]{inputenc} 

\usepackage{geometry} 
\usepackage{graphicx} 

\usepackage{booktabs} 
\usepackage{array} 
\usepackage{paralist} 
\usepackage{verbatim} 
\usepackage{subfig} 

\usepackage{fancyhdr} 
\usepackage{sectsty}
\allsectionsfont{\sffamily\mdseries\upshape} 

\usepackage[nottoc,notlof,notlot]{tocbibind} 
\usepackage[titles,subfigure]{tocloft} 

\title{Free will and quantum mechanics}
\author{Antonio Di Lorenzo\\
Instituto de F\'{\i}sica, Universidade Federal de Uberl\^{a}ndia,\\
 38400-902 Uberl\^{a}ndia, Minas Gerais, Brazil
}
\date{05/05/2011} 

\begin{document}
\maketitle
\begin{abstract}
A simple example is provided showing that violation of free will allows to reproduce the quantum mechanical predictions, 
and that the Clauser-Horne parameter can take the maximum value 4 for a proper choice.   
\end{abstract}
\begin{figure}
\centering
\includegraphics[width=4in]{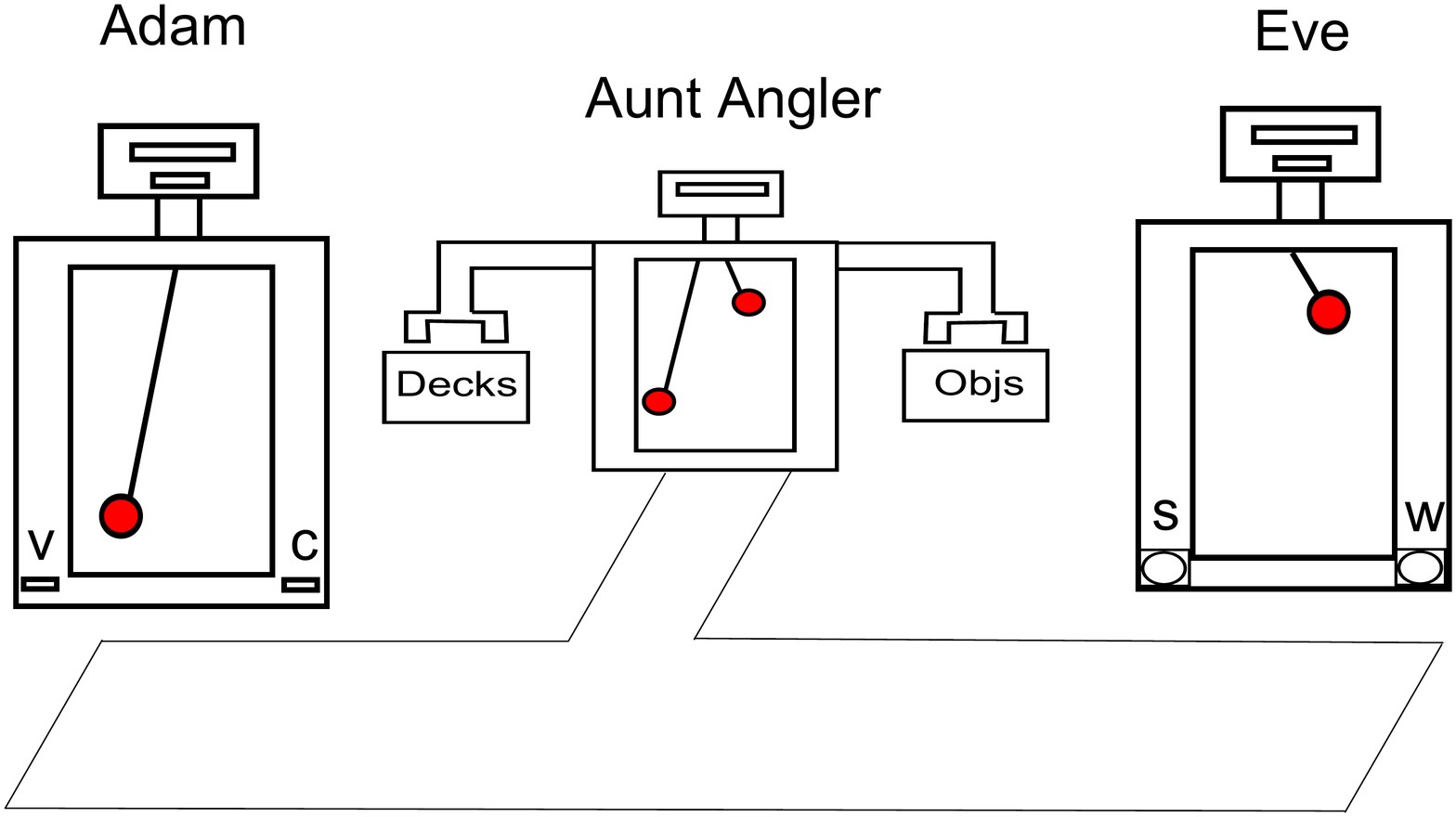}
\caption{\label{fig:setup}A sketch of the setup.}
\end{figure}

Quantum mechanics displays correlations which cannot be reproduced by local deterministic models \cite{Bell1964,Clauser1969,Clauser1974}. 
Some special classes of local stochastic models are also incompatible with quantum mechanics \cite{Leggett2003}. 
It has been argued that these conclusions rely on the hypothesis of free will on the part of the experimenters choosing 
the observables to be measured. While this claim is most probably correct, no actual example has been presented so far. 
In this Letter I report an instance of violation of Bell inequality which relies on classical objects and negation of free will. 
I have witnessed this violation in the laboratory of a programmer who wants to stay anonymous, so I will refer to him 
by the first letter of his name, G. 
Upon entering G's laboratory, I was greeted by two automata, whose names I soon learned to be Adam and Eve. 
They were sitting at the opposite ends of a long table, in the middle of which there was a third automaton, not endowed 
with speech, however. I believe Adam and Eve referred to her as ``Aunt Angler". 
Before the experiment started, I examined the automata, and discovered that Eve's mechanisms were connected to a pendulum. 
I found an identical pendulum inside Aunt Angler, which I soon realized was perfectly synchronized with Eve's. Inside 
Aunt Angler I found a second pendulum synchronized with an identical pendulum that was inside Adam. The period of Adam's pendulum was 
four times the period of Eve's. See Fig. \ref{fig:setup} for a sketch. 

The experiment started. Aunt Angler chose a card, which she sent to Adam, and an object, which she sent to Eve. The card and the object 
took the same time to get to Adam and Eve, and I found that this interval was exactly equal to the period of Adam's pendulum. 
Adam was so shaped that he could insert the card either into a hole where a device would determine the color of the card, 
which could be red ($R$) or black ($B$), or into another hole where a black and white scanner would determine the value, 
which could be a king ($K$) or a queen ($Q$). 
The object sent to Eve could be either a sphere ($S$) or a cube ($C$), sometimes massive, hence heavy ($H$), sometimes hollow, hence 
light ($L$).  Eve would examine the object she received either by weighting it or by determining its shape. 
I could not make out the complicated gears inside each automaton, but after observing the pendulums inside Aunt Angler 
I came to the following conclusions: Aunt Angler would pick an object from a mound where 
1/4 were Heavy Spheres, 1/4 Light Spheres, 1/4 Heavy Cubes, and 1/4 Light Cubes. 
There were four decks of cards. 
When both Adam's and Eve's pendulums were to the left, Aunt Angler would pick a card from deck number one ($D_1$) if the object she picked up 
was heavy, and from deck number two otherwise. 
I found out that $D_1$  had a fraction $(1-\sqrt{2}/2)/4$ of its cards made of Black Kings, as many Black Queens, while it had
$(1+\sqrt{2}/2)/4$ Red Queens and as many Red Kings. 
$D_2$, instead, had the opposite distribution, namely $(1-\sqrt{2}/2)/4$ of its cards were Red Kings, as many were Red Queens, and 
the remaining $(1+\sqrt{2}/2)/2$ cards were half Black Kings and half Black Queens. 
If Adam's pendulum was to the left and Eve's to the right, Aunt Angler would pick a card from $D_1$ whenever the object she had picked 
was a sphere, and from $D_2$ whenever it was a cube. 
Analogously, if Adam's pendulum was to the right, Aunt Angler would choose $D_3$ or $D_4$ depending on the position of Eve's pendulum 
and the object she had extracted from the mound. 
I will not bother further my readers, and resume the decision making process in Table \ref{table:dec}. 
\begin{table}
\begin{center}
\begin{tabular}{lll|c}
A&E&Object&Deck\\
\hline
left&left&Heavy&1\\
left&left&Light&2\\
left&right&Sphere&1\\
left&right&Cube&2\\
right&left&Heavy&3\\
right&left&Light&4\\
right&right&Sphere&4\\
right&right&Cube&3
\end{tabular}
\end{center}
\caption{\label{table:dec}The decision making algorithm of Aunt Angler.}
\end{table}
The composition of the decks is given instead in Table \ref{table:comp}. 
\begin{table}
\begin{center}
\begin{tabular}{l|cccc}
&BK&BQ&RK&RQ\\
\hline
$D_1$&
$\frac{1}{4}\left(1-\frac{\sqrt{2}}{2}\right)$&
$\frac{1}{4}\left(1-\frac{\sqrt{2}}{2}\right)$&
$\frac{1}{4}\left(1+\frac{\sqrt{2}}{2}\right)$&
$\frac{1}{4}\left(1+\frac{\sqrt{2}}{2}\right)$\\
$D_2$&$\frac{1}{4}\left(1+\frac{\sqrt{2}}{2}\right)$&$\frac{1}{4}\left(1+\frac{\sqrt{2}}{2}\right)$&$\frac{1}{4}\left(1-\frac{\sqrt{2}}{2}\right)$&$\frac{1}{4}\left(1-\frac{\sqrt{2}}{2}\right)$\\
$D_3$&$\frac{1}{4}\left(1-\frac{\sqrt{2}}{2}\right)$&$\frac{1}{4}\left(1+\frac{\sqrt{2}}{2}\right)$&$\frac{1}{4}\left(1-\frac{\sqrt{2}}{2}\right)$&$\frac{1}{4}\left(1+\frac{\sqrt{2}}{2}\right)$\\
$D_4$&$\frac{1}{4}\left(1+\frac{\sqrt{2}}{2}\right)$&$\frac{1}{4}\left(1-\frac{\sqrt{2}}{2}\right)$&$\frac{1}{4}\left(1+\frac{\sqrt{2}}{2}\right)$&$\frac{1}{4}\left(1-\frac{\sqrt{2}}{2}\right)$
\end{tabular}
\end{center}
\caption{\label{table:comp}The composition of the four decks.}
\end{table}
I also noticed that Adam and Eve were programmed so that when Adam's pendulum was to the left, he would choose to determine 
the color ($c$); otherwise he would choose to determine the value ($v$); 
analogously, Eve would always choose to measure the weight $w$ of the object 
when her pendulum was to the left, and the shape $s$ when it was to the right. 
I associated the values $+1$ to the outcomes $B,K,H,S$ and the values 
$-1$ to the remaining outcomes $R,Q,L,C$. 
I found out that the Clauser-Horne inequality is violated, since the parameter takes the value 
\begin{equation}
\mathcal{E}=|C(c,w)+C(c,s)+C(v,w)-C(v,s)|=2\sqrt{2} , 
\end{equation}
where $C$ indicates  the correlator, 
\begin{align}
C(a,b)=\sum_{o_a,o_b=\pm 1} o_a o_b P(o_a,o_b|a,b). 
\end{align}
The joint probability is indicated by the symbol $P(o_a,o_b|a,b)$, $a$ can be any $c$ or $v$ and $b$ can be $w$ or $s$. 
Finally, I took the liberty to substitute the four decks with other four, whose composition is given in Table \ref{table:comp2}. 
\begin{table}
\begin{center}
\begin{tabular}{l|cccc}
&BK&BQ&RK&RQ\\
\hline
$D_1$&$0$&$0$&$1/2$&$1/2$\\
$D_2$&$1/2$&$1/2$&$0$&$0$\\
$D_3$&$0$&$1/2$&$0$&$1/2$\\
$D_4$&$1/2$&$0$&$1/2$&$0$
\end{tabular}
\end{center}
\caption{\label{table:comp2}The composition of the four new decks.}
\end{table}
Then the Clauser-Horne parameter takes the maximum value $\mathcal{E}=4$. 

As a funny side note, G programmed Adam and Eve with an astounding artificial intelligence. When I asked them how they chose 
which measurement to perform, they told me it was out of free will, and when I told them there was a pendulum inside them that 
would determine their decisions, they would dismiss this thought as far-fetched.  
Now I have to excuse myself from my readers, since G is asking me to switch some polarizers in his laboratory.


\begin{thebibliography}{9}

\bibitem{Bell1964}
J.~S. Bell, Physics \textbf{1}, 195 (1964).

\bibitem{Clauser1969}
J.~F. Clauser, M.~A. Horne, A.~Shimony, and R.~A. Holt, Phys. Rev. Lett.
  \textbf{23}, 880 (1969).

\bibitem{Clauser1974}
J.~F. Clauser and M.~A. Horne, Phys. Rev. D \textbf{10}, 526 (1974).

\bibitem{Leggett2003}
A.~J. Leggett, Found. Phys. \textbf{33}, 1469 (2003).

\end{thebibliography}
\end{document}